\documentclass[sn-mathphys,Numbered]{sn-jnl}
\usepackage{graphicx}
\usepackage{multirow}
\usepackage{amsmath,amssymb,amsfonts}
\usepackage{amsthm}
\usepackage{mathrsfs}
\usepackage[title]{appendix}
\usepackage{xcolor}
\usepackage{textcomp}
\usepackage{manyfoot}
\usepackage{booktabs}
\usepackage{algorithm}
\usepackage{algorithmicx}
\usepackage{algpseudocode}
\usepackage{listings}
\usepackage{epigraph}
\usepackage{hyperref}
\usepackage{url}
\theoremstyle{thmstyleone}
\newtheorem{theorem}{Theorem}
\newtheorem{proposition}[theorem]{Proposition}
\theoremstyle{thmstyletwo}
\newtheorem{example}{Example}
\newtheorem{remark}{Remark}

\theoremstyle{thmstylethree}

\begin{document}

\title[An RSA Cryptosystem over a Halidon Group Ring of a Dihedral Group ]{An RSA Cryptosystem over a Halidon Group Ring of a Dihedral Group }

\author*[1]{\fnm{A.} \sur{Telveenus}}\email{t.fernandezantony@kingston.ac.uk, telveenusa@gmail.com}

\affil*[1]{\orgdiv{International Study Centre}, \orgname{Kingston University}, \orgaddress{\street{Kingston Hill Campus}, \city{London}, \postcode{KT2 7LB},  \country{UK}}  \orgdiv{and Former HoD Mathematics, Fatima Mata National College}, \orgname{University of Kerala}, \orgaddress{\street{Kollam}, \city{Kerala},  \country{India}} }

\abstract{The article explores the creation of a cryptosystem using a halidon group ring of a dihedral group. Due to the non-abelian nature of the group, constructing the cryptosystem is more challenging compared to an abelian group. The logic used to develop a decryption programme was also quite complex.
}



\maketitle
\epigraph{What is circular is eternal! \\ What is eternal is circular!}{\textit{Philip J Davis}}

\section{Introduction}
The development of asymmetric cryptography, also known as public-key cryptography, is considered the most significant and perhaps the only true revolution in the history of cryptography. A significant portion of the theory behind public-key cryptosystems relies on number theory(\cite{ws}). RSA is a widely used public-key cryptosystem for secure data transmission. It was first publicly described in 1977 by Ron Rivest, Adi Shamir, and Leonard Adleman. An equivalent system had been secretly developed in 1973 by Clifford Cocks at the British signals intelligence agency, GCHQ. \\
In 1940, the famous celebrated mathematician Graham Higman published a theorem \cite{gh},\cite{gk} in group algebra which is valid only for a field or an integral domain with some specific conditions. In 1999, the author noticed that this theorem can be extended to a rich class of rings called halidon rings\cite{at}.  \\

A primitive $m^{th}$ root of unity in a ring with unit element is completely different from that of in a field, because of the presence of nonzero zero divisors.  So we need a separate definition for a primitive $m^{th}$ root of unity. An element $\omega $ in a ring $R$ is called a  \textit{primitive} $m^{th}$ root if $m$ is the least positive integer such that  $\omega^{m}=1$ and
\begin{eqnarray*}
\sum_{r=0}^{m-1} \omega^{r(i-j)}&=& m, \quad  i= j (\ mod \ m )\\ &=& 0, \quad  i\neq j (\ mod \ m ). \end{eqnarray*} \\
More explicitly,
\begin{eqnarray*}
1+ \omega^{r}+(\omega^{r})^{2}+(\omega^{r})^{3}+(\omega^{r})^{4}+......+(\omega^{r})^{m-1}&=& m, \quad  r=0 \\ &=& 0, \quad 0<r\leq m-1. \end{eqnarray*} \\
A ring $R$ with unity is called a \textit{halidon} ring with index $m$ if there is a  primitive $m^{th}$ root of unity and $m$ is invertible in $R$. The ring of integers is a halidon ring with index $ m=1$ and $\omega=1$. The halidon ring with index $1$ is usually called a \textit{trivial} halidon ring. 
\section{Preliminary results}
In this section, we shall state the results essential for the construction the cryptosytem over a halidon group ring of a dihedral group only. The readers who are interested in the properties of halidon rings and their applications, can refer to \cite{at}, \cite{ath}, \cite{ath1} and \cite{ath2}. \\
Let $D_{m} =<a,b|a^{m}=1, b^{2}=1, abab =1>$ 
be the \textit{dihedral group} of order $2m$ ,( here $s= ab$ and $t= b$ ) (\cite{rdc} ) \\
This non abelian group can be rewritten as 
$D_{m} = \{ a^{i-1}b^{j-1} | i= 1,2,...., m \ and \ j= 1,2 \}$ where $ba=a^{m-1}b.$
Any element in $RD_{m}$ can be taken as \begin{equation}
u=\sum_{i=1}^{m}\alpha_{i}a^{i-1} + \sum_{i=1}^{m}\alpha_{m+i}a^{i-1}b
\end{equation}
where each $\alpha_{k} \in R$ for $k= 1,2,....,m,m+1,.....2m$. 
\begin{theorem}\cite{pjd} \cite{at}
$RD_{m} \cong V$ , as $R$ -algebras,  where $V$ is a subalgebra of the algebra
of matrices, $Mat_{2m} (R)$, of order $2m$ over $R$ and therefore $U(RD_{2m}) \cong U(V)$.
\end{theorem}
The key isomorphism in the above theorem is $u\longrightarrow A_{u}$ where \\
$$A_{u}=\left(
         \begin{array}{cc}
           Ciruc(\alpha_{1}, \ \alpha_{m}, \ \alpha_{m-1}, ..., \alpha_{2}) & Ciruc(\alpha_{m+1}, \ \alpha_{m+2}, \ \alpha_{m+3}, ..., \alpha_{2m}) \\
           Ciruc(\alpha_{m+1}, \ \alpha_{2m}, \ \alpha_{2m-1}, ..., \alpha_{m+2}) & Ciruc(\alpha_{1}, \ \alpha_{2}, \ \alpha_{3}, ..., \alpha_{m}) \\
         \end{array}
       \right)
.$$ To know more about circulants, read \cite{pjd}.
\begin{theorem} \label{rd5}
Let $$u=\sum_{i=1}^{m}\alpha_{i}a^{i-1} + \sum_{i=1}^{m}\alpha_{m+i}a^{i-1}b$$ be an invertible element in $RD_{m}$ and let  $$u^{-1}=\sum_{i=1}^{m}\beta_{i}a^{i-1} + \sum_{i=1}^{m}\beta_{m+i}a^{i-1}b.$$ Then there exist $l_{r}$, $s_{r}$ $\in R$; $i=1,2,3....m$ such that \begin{equation} 
\beta_{i}=\frac{1}{m} \sum _{r=1}^{m} l_{r} (\omega^{i-1})^{r-1}; i=1,2,3...,m
\end{equation}
and
\begin{equation} 
\beta_{m+i}=\frac{1}{m} \sum _{r=1}^{m} s_{r} (\omega^{i-1})^{r-1}; i=1,2,3...,m.
\end{equation}
\end{theorem}
From the proof of the above theorem (\cite{at}), $$l_{r}=\eta_{r}(\lambda_{r}\eta_{r}-\gamma_{r}\delta_{r})^{-1}$$ and $$s_{r}=-\delta_{r}(\lambda_{r}\eta_{r}-\gamma_{r}\delta_{r})^{-1},$$ 
where $$\lambda_{i}=\sum_{r=1}^{m} \alpha_{m-r+2}(\omega^{(i-1)})^{(r-1)},$$
$$\gamma_{i}=\sum_{r=1}^{m} \alpha_{m+r}(\omega^{(i-1)})^{(r-1)},$$
$$\delta_{i}=\sum_{r=1}^{m} \alpha_{m+(m-r+2)}(\omega^{(i-1)})^{(r-1)},$$
and 
$$\eta_{i}=\sum_{r=1}^{m} \alpha_{r}(\omega^{(i-1)})^{(r-1)},$$
where $\omega \in R$ is a primitive $m^{th}$ root of $1$ and $m-r+2$ is being taken reduction
modulo $m$ and zero is treated as $m$ and $i=1,2,.....,m.$
\begin{theorem} \label{rd6}
Let $l_{i}, s_{i} ,\lambda_{i}, \gamma_{i}, \delta_{i}, \eta_{i} \in R ; i=1,2,.....m$ be as in Theorem \ref{rd5}.
Then
\begin{enumerate}
  \item $\lambda_{m-i+2}=\eta_{i}$
  \item $\gamma_{m-i+2}=\delta_{i}$
  \item $l_{i}=\lambda_{m-i+2}(\lambda_{i}\lambda_{m-i+2}-\gamma_{i}\gamma_{m-i+2})^{-1}$
  \item $s_{i}=-\gamma_{m-i+2}(\lambda_{i}\lambda_{m-i+2}-\gamma_{i}\gamma_{m-i+2})^{-1}$
\end{enumerate}

\end{theorem}
For programme 1 use programme 1 in \cite{ath2}. \\
The following decryption programme has been developed based on the theorems \ref{rd5} and \ref{rd6}. A reader can assess the level of difficulty of the programme logic used in the computer programme-2 very easily. 
 
 Computer programme-2 \\
 \begin{verbatim}
#include<iostream>
#include<cmath>
// lambda c[r]
// Gamma c1[r]
// u[r]^-1 b[r]
//l[r]  d[r]
//s[r]  d1[r]
using namespace std;
int main() {
	cout << "To check whether an element in Z(n)G;" <<
		"G is a dihedral group of order 2m" << "has a multiplicative inverse or not" << endl;
	long long	int a[1000], b[1000], b1[1000], c[1000], c1[1000], d[1000],d1[1000], 
                e[1000], w1[1000], u[1000],f[1000],g[1000], m = 1,
		t = 0, x = 1, s = 0, s1 = 0, l = 0, l1 = 0, m1 = 1, hcf = 1,
		n = 1, i = 1, k = 0, k1 = 0, q = 1, p = 1, r = 1, r1 = 1, r2 = 1, r3 = 1, w = 1;
	cout << "Enter n =" << endl;
	cin >> n;
	cout << "Enter index m =" << endl;
	cin >> m;
	cout << "Enter  m^(-1) =" << endl;
	cin >> m1;
	cout << "Enter primitive m th root w =" << endl;
	cin >> w;
	for (i = 1; i < m; ++i) { c1[i] = 0; }
	w1[0] = 1;
	for (i = 1; i < m; ++i)
	{
		//w1[i] = ((long long int)pow(w, i)) % n;
		w1[i] = p * w % n; p = w1[i];
		cout << "w1[" << i << "]" << w1[i] << endl;
	}
	for (int i = 1; i < 2 * m + 1; ++i) {
		cout << "Enter a[" << i << "]=" << endl;
		cin >> a[i];
	}
	a[0] = a[m];
	for (int r = 1; r < m + 1; ++r) {
		for (int j = 1; j < m + 1; ++j)
		{
			l = (m - j + 2) % m;
			x = ((j - 1) * (r - 1)) % m;
			k = k + (a[l] * w1[x]) % n; k = k % n;
			cout << "k=" << k << endl;
		} c[r] = k; cout << "lambda[" << r << "]=" << c[r] << endl;
		k = 0;
	}
	for (int r1 = 1; r1 < m + 1; ++r1) {
		for (int j1 = 1; j1 < m + 1; ++j1) {
			int l1 = m + j1;
			cout << "l1=" << l1 << endl;
			int x1 = ((j1 - 1) * (r1 - 1)) % m;
			k1 = k1 + (a[l1] * w1[x1]) % n; k1 = k1 % n; cout << "k1=" << k1 << endl;
		}c1[m + r1] = k1; cout << "gamma[" << m + r1 << "]=" << c1[m + r1] << endl; k1 = 0;
	}
	c[0] = c[m];
	for (r2 = 1; r2 < m + 1; ++r2) {
		if (r2 == 2) {
			u[r2] = (c[r2] * c[(m - r2 + 2) % m] - c1[m + r2] * c1[2 * m]) % n; 
       if (u[r2] < 0) { u[r2] = u[r2] + n; }
		}
		else {
			u[r2] = (c[r2] * c[(m - r2 + 2) % m] - c1[m + r2] * c1[m + (m - r2 + 2) % m]) % n; 
       if (u[r2] < 0) { u[r2] = u[r2] + n; }
		}
	}


	for (r3 = 1; r3 < m + 1; ++r3) {
		for (int i = 1; i <= n; ++i) {
			if (u[r3] % i == 0 && n % i == 0) {
				hcf = i;
			}
		}
		if (hcf == 1) {
			cout << "u[" << r3 << "] ="<< u[r3]<<" is a unit" << endl;
		}
		else {
			cout << "u[" << r3 <<
				"]=" <<u[r3]<< " is a not unit.So there is no multiplicative inverse." <<
				endl; t = 1;
		}
	}
	for (r = 1; r < m + 1; ++r) {
		for (int i = 1; i <= n; ++i) {
			e[r] = (u[r] * i) % n;
			if (e[r] == 1) {
				b[r] = i;
				cout << " The inverse of u[" << r << "] is " << b[r] << endl;
			}
		}
	}

	f[0] = f[m];
	for (r = 1; r < m + 1; ++r) {
		f[r] = (c[(m - r + 2)%m] * b[r])%n;
		cout << "f[" << r << "]= " << f[r] << endl;
	}
	for (int r = 1; r < m + 1; ++r) {
		for (int j = 1; j < m + 1; ++j)
		{
			x = ((j - 1) * (r - 1)) % m;  //cout << "x= " << x << endl;
			k = k + (m1 * f[j] * w1[x]) % n; k = k % n;
			//cout << "k=" << k << endl;
		}
		d[r] = k; cout << "d[" << r << "]=" << d[r] << endl;
		k = 0;
	}
	g[0] = g[m];
	for (r = 1; r < m + 1; ++r)
	{
		x = 2 * m;
	int	y = m + (m - r + 2) % m;
		if (r == 2) { g[r] = (-c1[x] * b[r]) % n; if (g[r] < 0) { g[r] = g[r] + n; } }
		else {
			g[r] = (-c1[y] * b[r])%n; if (g[r] < 0) { g[r] = g[r] + n; }
		}cout << "g[" << r << "]= " << g[r] << endl;
	}
	for (int r = 1; r < m + 1; ++r) {
		for (int j = 1; j < m + 1; ++j)
		{
			x = ((j - 1) * (r - 1)) % m;  //cout << "x= " << x << endl;
			k = k + (m1 * g[j] * w1[x]) % n; k = k % n;
			//cout << "k=" << k << endl;
		}
		d1[r] = k; cout << "d1[" << r << "]=" << d1[r] << endl;
		k = 0;
	}
	if (t == 1) {
		s = m;
	mylabel2:
		cout << a[m - s + 1] << "g^(" << m - s << ") + ";
		s--;
		if (s > 0) goto mylabel2; cout <<
			"has no multiplicative inverse." << endl;
	}
	else {
		cout << "The inverse of ";
		for (i = 1; i < m + 1; ++i) {
			cout << a[i] << "a^(" << i - 1 << ")+";
		}
		for (i = m + 1; i < 2 * m + 1; ++i) {
			cout << a[i] << "a^(" << i - m - 1 << ")b+";
		}cout << "is  " << endl;
		for (i = 1; i < m + 1; ++i) {
			cout << d[i] << "a^(" << i - 1 << ")+";

		}

		for (i = m + 1; i < 2 * m + 1; ++i) {
			cout << d1[i-m] << "a^(" << i - m - 1 << ")b+";


		}
	}
	return 0;
}

 \end{verbatim}
 
\begin{proposition} \cite{jl} \label{rd8}
Let $n=p_{i}^{e_{1}}p_{2}^{e_{2}}....p_{k}^{e_{k}}$ be the standard form of the integer $n$ and let d,e satisfy $ed\equiv 1 \ mod \ \phi(n)$. Then for all integer $x$,
$$x^{ed}\equiv x \ mod \ n.$$ Therefore, if $c=x^{e} \mod \ n$, we have $x \equiv c^{d}\ mod \ n. $
\end{proposition}
\section{An RSA Cryptosystem over a halidon ring of a dihedral group}
Let $2m$ be the length of the message including the blank spaces between the words. If the message has a length more than $2m$, we can split the message into blocks with lengths less than $2m$. For a message of length less than $2m$, we can add blank spaces after the period to make it a message with length $2m$. Choose  two large prime numbers such that $p_{i}=mt_{1}+1$ where $i=1,2$ and  $t_{i}$'s are relatively prime. Let $\omega$ be a primitive $m^{th}$ root of unity in $\mathbb{Z}_{n}$, where $n=p_{1}p_{2}.$ We know that $\mathbb{Z}_{n}$ is a halidon ring with maximum index $gcd(p_{1}-1,p_{2}-1)$(\cite{ath2} and since $m|p_{1}-1,p_{2}-1,$ $\mathbb{Z}_{n}$ is also a halidon ring with index $m$(\cite{ath2}).

Here we are considering a cryptosystem based on modulo $n$. The following table gives numbers and the corresponding symbols.
\begin{center}
\begin{tabular}{|c|c|}

 \hline
  Numbers assigned & Symbols \\
  \hline
  1 to 9 & 1 to 9 \\
  10 to 35  & A to Z \\
  36 & blank space \\
  37 & colon \\
  38 & period \\
  39 & hyphen \\
  40 & + \\
  41 & 0 \\
  42 & comma \\
\hline
\end{tabular}
\end{center}
\begin{remark}
The numbers chosen to assign symbols must be done in such a way that $(\lambda_{i}\lambda_{m-i+2}-\gamma_{i}\gamma_{m-i+2})$ must be invertible in $R$. This task is also a bit challenging. 
\end{remark}

In this cryptosystem, there are two stages. In stage 1, we shall compute the value of $\omega$ which Bob keeps secret and in stage 2, we shall decrypt the message sent by Bob. \\
\textbf{Stage 1-RSA}\\
\textbf{Cryptosystem setup}
\begin{enumerate}
  \item Alice chooses two large primes $p_{1}, p_{2}$ and calculates $n=p_{1}p_{2}$ and $\phi(n)=(p_{1}-1)(p_{2}-1)$.
  \item Alice chooses an $e$ so that $gcd(e,\phi(n))=1$.
  \item Alice calculates $d$ with property $ed\equiv \ 1 mod \ \phi(n)$.
  \item Alice makes $n$ and $e$ public and keeps the rest secret$(p_{1},p_{2}).$
\end{enumerate}
\textbf{Cryptosystem Encryption}(Programme-1)
\begin{enumerate}
  \item Bob looks up Alice's $n$ and $e$ .
  \item Bob chooses an arbitrary $\omega\ mod \ n$ and kept secret.
  \item Bob sends $c \equiv \omega^{e} \ mod \ n$ to Alice.
\end{enumerate}
\textbf{Cryptosystem Decryption}(Proposition \ref{rd8})
\begin{enumerate}
  \item Alice receives $c$ from Bob.
  \item Alice computes $\omega\equiv \ c^{d}mod \ n$.
\end{enumerate}
\textbf{Stage 2-Dihedral group}\\
\textbf{Cryptosystem setup}
\begin{enumerate}
  \item Alice chooses programme -2  as the encryption key.
  \item Alice chooses programme-2  as the decryption key where output of the encryption key is used as input. 
\end{enumerate}
\textbf{Cryptosystem Encryption}(Programme-2)
\begin{enumerate}
  \item Bob looks up Alice's encryption key.
  \item Bob writes his message $x$.
  \item Bob computes output $y$ using encryption key with his chosen $\omega$.
  \item Bob sends $y$ to Alice.
\end{enumerate}

\textbf{Cryptosystem Decryption}(Programme-2 using the output as input)
\begin{enumerate}
  \item Alice receives $y$ from Bob.
  \item Alice computes $x$ with $\omega$ calculated in stage 1 using the decryption key.
\end{enumerate}
The above cryptosystem is an \textbf{asymmetric cryptosystem} as Alice and Bob share different information. For the practical application, we must choose very large prime numbers (more than 300 digits) so that the calculation of $\phi(n)$ must be very difficult and the probability of choosing the primitive $m^{th}$root of unity $\omega$  should tends to zero. 
\begin{example}
\textbf{Stage 1}\\
The length of the message has been fixed as $2m=100$.
Alice chooses two primes $p_{1}=151$and $p_{2}=701$ and calculates $n=105851$ and $\phi(n)=105000$.\\
Alice chooses an $e=65537$ so that $gcd(e,\phi(n))=1$.\\
Alice calculates $d=48473$ with property $ed\equiv \ 1  \ mod \ \phi(n)$.
Alice shared Bob $n=105851$ and $e=65537$ and rest kept secret. \\
Bob looks up Alice's $n=105851$ and $e=65537$. \\
Bob chooses an arbitrary $50^{th}$ root of unity $\omega\ mod \ n$(there are $\phi(50)^{2}=400$ $\omega$'s possible and they can be found by running the programme $1$ and kept secret. \\
Bob sends $c \equiv \omega^{e} \ mod \ n \equiv 104726$ \ mod 105851 to Alice.
Alice receives $c$ from Bob.\\
Alice computes $\omega\equiv \ c^{d}mod \ n\equiv 37199 \ mod \ 104726$.\\
\textbf{Stage 2} \\

Alice shared Bob the encryption key PROGRAMME-2 and  $n=104726.$
Suppose Bob sends the following secret message to Alice. \\
\begin{center} NAME: XYACDX AGE: 67 BLOOD GROUP:G+ EYE SIGHT: BLIND ADDRESS: XL40, ROMA CANCER PATIENT. \end{center}
The length of the message is 89 and to make it 100 we need to add 11 blank spaces. This can be translated into a 100 component vector \\

 $ \left(
                                   \begin{array}{c}
23 \ 10 \ 22 \ 14 \ 37 \ 36 \ 33 \ 34 \ 10 \ 12 \ 13 \ 33 \ 36 \\  
10 \ 16 \ 14 \ 37 \ 36 \ 6 \ 7 \ 36 \ 11 \ 21 \ 24 \ 4 \ 13 \ 36 \\  
16 \ 27 \ 24 \ 30 \ 25 \ 37 \ 36 \ 16 \ 40 \ 36 \\            
14 \ 34 \ 14 \ 36 \ 28 \ 18 \ 16 \ 17 \ 29 \ 37 \ 36 \\         
11 \ 21 \ 18 \ 23 \ 13 \ 36 \ 10 \ 13 \ 13 \ 27 \ 14 \ 28 \ 28 \ 37 \ 36 \\  
33 \ 21 \ 4 \ 41 \ 42 \ 36 \ 27 \ 24 \ 22 \ 10 \ 36 \\      
12 \ 10 \ 23 \ 12 \ 14 \ 27 \ 36 \ 25 \ 10 \ 29 \ 18 \ 14 \ 23 \ 29 \ 38 \\  
  36 \ 36 \ 36 \ 36 \ 36 \ 36 \ 36 \ 36 \ 36 \ 36 \ 36

                                   \end{array}
                                 \right)$

\noindent
Bob chooses a primitive $50^{th}$ roots of unity $\omega$ in stage 1 using  programme 1 and kept secret in the halidon ring $\mathbb{Z}_{105851}$ with index 50.
Applying programme 2 to the plain text to get the following cipher text using the chosen value of $\omega$:  \\

$ \left(
                                   \begin{array}{c}
                                   30250 \ 3997 \ 9918 \ 100174 \ 4967 \ 60850 \ 69603 \ 65833 \ 69970 \ 86837 \\
                                   56329 \ 84304 \ 87528 \ 105689 \ 50220 \ 37821 \ 95006 \ 88206 \ 50634 \ 56876 \\
                                   96029 \ 12250 \ 5412 \ 92277 \ 81732 \ 97464 \ 88405 \ 83966 \ 90468 \ 47910 \\
                                   96334 \ 58983 \ 53641 \ 28780 \ 81547 \ 86900 \ 39702 \ 55100 \ 86014 \ 58257 \\
                                   86271 \ 84815 \ 94860 \ 52442 \ 70408 \ 52434 \ 66586 \ 4211 \ 51571 \ 56249 \\
                                   26550 \ 22598 \ 45775 \ 88780 \ 53508 \ 4039 \ 20356 \ 54096 \ 97408 \ 31956  \\
                                   74800 \ 74474 \ 100196 \ 69161 \ 70858 \ 80036 \ 95190 \ 7660 \ 58416 \ 42076 \\
                                   10798 \ 10907 \ 24694 \ 2436 \ 50661 \ 80202 \ 19348 \ 97383 \ 31692 \ 67941  \\
                                   8843 \ 97849 \ 63925 \ 67146 \ 28787 \ 31773 \ 83630 \ 100041 \ 13855 \ 5141 \\
                                   99471 \ 55282 \ 2137 \ 56462 \ 24205 \ 99031 \ 94986 \ 44166 \ 101479 \ 35249
                                     \end{array}
                                 \right) $

\noindent

The readers can check the above results by copying the programmes and paste in Visual Studio 2022 c++ projects. \\
\begin{verbatim}
The output of programme-2
The inverse of 23a^(0)+10a^(1)+22a^(2)+14a^(3)+37a^(4)+36a^(5)+33a^(6)+
34a^(7)+10a^(8)+12a^(9)+13a^(10)+33a^(11)+36a^(12)+10a^(13)+16a^(14)+
14a^(15)+37a^(16)+36a^(17)+6a^(18)+7a^(19)+36a^(20)+11a^(21)+21a^(22)+
24a^(23)+4a^(24)+13a^(25)+36a^(26)+16a^(27)+27a^(28)+24a^(29)+30a^(30)+
25a^(31)+37a^(32)+36a^(33)+16a^(34)+40a^(35)+36a^(36)+14a^(37)+34a^(38)+
14a^(39)+36a^(40)+28a^(41)+18a^(42)+16a^(43)+17a^(44)+29a^(45)+37a^(46)+
36a^(47)+11a^(48)+21a^(49)+18a^(0)b+23a^(1)b+13a^(2)b+36a^(3)b+10a^(4)b+
13a^(5)b+13a^(6)b+27a^(7)b+14a^(8)b+28a^(9)b+28a^(10)b+37a^(11)b+
36a^(12)b+33a^(13)b+21a^(14)b+4a^(15)b+41a^(16)b+42a^(17)b+
36a^(18)b+27a^(19)b+24a^(20)b+22a^(21)b+10a^(22)b+36a^(23)b+12a^(24)b+
10a^(25)b+23a^(26)b+12a^(27)b+14a^(28)b+27a^(29)b+36a^(30)b+25a^(31)b+
10a^(32)b+29a^(33)b+18a^(34)b+14a^(35)b+23a^(36)b+29a^(37)b+38a^(38)b+
36a^(39)b+36a^(40)b+36a^(41)b+36a^(42)b+36a^(43)b+
36a^(44)b+36a^(45)b+36a^(46)b+36a^(47)b+36a^(48)b+36a^(49)b+is

30250a^(0)+3997a^(1)+9918a^(2)+100174a^(3)+4967a^(4)+60850a^(5)+
69603a^(6)+65833a^(7)+69970a^(8)+86837a^(9)+56329a^(10)+84304a^(11)+
87528a^(12)+105689a^(13)+50220a^(14)+37821a^(15)+95006a^(16)+
88206a^(17)+50634a^(18)+56876a^(19)+96029a^(20)+12250a^(21)+5412a^(22)+
92277a^(23)+81732a^(24)+97464a^(25)+88405a^(26)+83966a^(27)+
90468a^(28)+47910a^(29)+96334a^(30)+58983a^(31)+53641a^(32)+28780a^(33)+
81547a^(34)+86900a^(35)+39702a^(36)+55100a^(37)+86014a^(38)+
58257a^(39)+86271a^(40)+84815a^(41)+94860a^(42)+52442a^(43)+70408a^(44)+
52434a^(45)+66586a^(46)+4211a^(47)+51571a^(48)+56249a^(49)+
26550a^(0)b+22598a^(1)b+45775a^(2)b+88780a^(3)b+53508a^(4)b+
4039a^(5)b+20356a^(6)b+54096a^(7)b+97408a^(8)b+31956a^(9)b+
74800a^(10)b+74474a^(11)b+100196a^(12)b+69161a^(13)b+70858a^(14)b+
80036a^(15)b+95190a^(16)b+7660a^(17)b+58416a^(18)b+42076a^(19)b+
10798a^(20)b+10907a^(21)b+24694a^(22)b+2436a^(23)b+50661a^(24)b+
80202a^(25)b+19348a^(26)b+97383a^(27)b+31692a^(28)b+67941a^(29)b+
8843a^(30)b+97849a^(31)b+63925a^(32)b+67146a^(33)b+28787a^(34)b+
31773a^(35)b+83630a^(36)b+100041a^(37)b+13855a^(38)b+5141a^(39)b+
99471a^(40)b+55282a^(41)b+2137a^(42)b+56462a^(43)b+24205a^(44)b+
99031a^(45)b+94986a^(46)b+44166a^(47)b+101479a^(48)b+35249a^(49)b+
C:\Users\DrTel\Documents\C++\Dihedral\x64\Debug\Dihedral.exe (process 40468) 
exited with code 0.
Press any key to close this window . . .
Neglect the last + which is unavoidable in the programme. 

\end{verbatim}

Alice receives the above cipher text and  she uses  $\omega=37199$ from stage 1. Applying programme 2 again Alice gets the original message back. Also, we can assign letters and numbers in $nP42$ ways which will also make the adversaries their job difficult. For messages with length more than $2m$, split the message into blocks with length less than $2m$.
\end{example}
\section{Conclusion}\label{sec13}
The new cryptosystem has been developed using  halidon group rings of a dihedral group.The system provides a
high-level security for communication between ordinary people or classified messages in government agencies. The level of security can be increased by utilising advanced computer technology and powerful codes to calculate the primitive $m^{th}$ root of unity for a very large value of $n$ where the calculation of $\phi(n)$ is difficult. There are scopes for the development of new cryptosystems based on dicyclic and other non-ablian groups.

\end{document}